\documentclass[%
nofootinbib,
superscriptaddress,
preprint,
 amsmath,amssymb,ams
 aps,
]{revtex4-2}

\usepackage{ulem}
\usepackage{bbm}
\usepackage{graphicx}
\usepackage{dcolumn}
\usepackage{bm}

\usepackage{multirow}
\usepackage{pgfplots}
\usepackage{tikz}
\usepackage{slashed}
\usepackage{braket}
\usepackage{cancel}
\usetikzlibrary{shapes.misc}
\tikzset{cross/.style={cross out, draw=black, minimum size=2*(#1-\pgflinewidth), inner sep=0pt, outer sep=0pt},
cross/.default={1pt}}
\begin{document}
\preprint{KEK-TH-2717}
\preprint{J-PARC-TH-0316}
\preprint{RIKEN-iTHEMS-Report-25}
\title{Pole-Expansion of Two-Hadron Imaginary-Time Correlation Function \\
-a new method of analysis for unstable states in lattice QCD-}
\date{\today}
\author{Wren Yamada}
\email{rento.yamada@riken.jp}
\affiliation{RIKEN iTHEMS, Wako, Saitama 351-0198, Japan}%

\author{Osamu Morimatsu}
\email{osamu@post.kek.jp}
\affiliation{Theory Center, Institute of Particle and Nuclear Studies (IPNS), High Energy Accelerator Research Organization (KEK), 1-1 Oho, Tsukuba, Ibaraki, 205-0801, Japan}%

\author{Toru Sato}
\email{tsato@rcnp.osaka-u.ac.jp}
\affiliation{Research Center for Nuclear Physics (RCNP), Osaka University, Ibaraki, Osaka 567-0047, Japan}%

\author{Koichi Yazaki}
\email{koichiyzk@yahoo.co.jp}
\affiliation{RIKEN iTHEMS, Wako, Saitama 351-0198, Japan}%

\begin{abstract}
We analyze the pole expansion of the two-hadron imaginary-time correlation function.
We first explain the general idea that the imaginary-time correlation function is expressed as a sum of the pole terms, the Mittag-Leffler expansion, in terms of the uniformization variable, which makes the S-matrix single-valued.
We then derive explicit expressions of the pole expansion for the single-channel ($\rho$ meson) and two-channel ($\Lambda(1405)$) examples and demonstrate that the pole expansion actually holds employing phenomenological models, the vector-dominance model for the $\rho$ meson and the chiral unitary model for $\Lambda(1405)$.
From this observation we propose the pole expansion as a method to extract information of unstable states such as masses and widths from the two-hadron imaginary-time correlation functions obtained by lattice QCD simulations.
\end{abstract}
\maketitle

One of the most unique phenomena in strong interaction physics is the existence of hadron resonances.
In the study of hadron resonances by lattice QCD \cite{Briceno:2017max,CP-PACS:2007wro,Feng:2010es,CS:2011vqf,Lang:2011mn,Dudek:2012xn,Alexandrou:2017mpi,Andersen:2018mau,Wilson:2015dqa,Dudek:2016cru,Dudek:2014qha,BaryonScatteringBaSc:2023zvt,BaryonScatteringBaSc:2023ori,Moir:2016srx}, the information of hadron interactions is extracted from the imaginary-time correlation function of two hadrons.
The method of extraction most often used is generically called the L\"{u}scher method \cite{Luscher:1985dn,Luscher:1986pf,Rummukainen:1995vs}, in which the energy shifts of two hadrons in a finite volume are related to the infinite-volume scattering amplitude by the L\"{u}scher quantization condition. 
HALQCD method is another method \cite{Ishii:2006ec,Aoki:2009ji}, where the Nambu-Bethe-Salpeter wave function is calculated and used to determine a potential between two hadrons.
Recently, lattice QCD has started to be applied to hadron resonances in the coupled-channel scatterings \cite{Dudek:2014qha,Wilson:2015dqa,Dudek:2016cru,BaryonScatteringBaSc:2023zvt,BaryonScatteringBaSc:2023ori,Moir:2016srx}. 
Even though these attempts are groundbreaking, there is still much room for improvement.
In Ref.\,\cite{BaryonScatteringBaSc:2023zvt,BaryonScatteringBaSc:2023ori} several steps are needed in order to obtain masses and widths of resonances from the imaginary-time correlation function and also some arbitrary assumptions are made in the parametrization of the $K$-matrix, which can be improved in our opinion.

In the present paper we analyze the pole expansion of the imaginary-time correlation function in general and derive explicit expressions of the pole expansion for single and coupled two-channels.
The method has been developed and applied to the study of  unstable states in coupled channels in Refs.\,\cite{HUMBLET1961529, Yamada:2020rpd, Yamada:2021azg, Yamada:2023wqp}, where unstable states refer to poles more general than resonances and will be hereafter used.
We neglect the left-hand cuts and three- or more-body channels and assume that the correlation function is a meromorphic function of an appropriate variable, the uniformization variable, which makes the correlation function single-valued.
Then, the correlation function is expanded as a sum of pole terms by virtue of the Mittag-Leffler theorem.
The obtained results directly relate the imaginary-time correlation functions with the masses and widths of unstable states since the pole expansion is parametrized by the pole positions and residues of unstable states.
We then demonstrate the pole expansion both for the single-channel ($\rho$ meson) and two-channel ($\Lambda(1405)$) examples.
From this observation we propose the pole expansion of the imaginary-time correlation as a method to extract information of unstable states such as masses and widths from the imaginary-time correlation functions obtained by lattice QCD simulations.

Consider the imaginary-time correlation function
\begin{align}
  C^{ij} (\tau) &= \bra{0} {\cal O}^i(\tau)  {\cal O}^{j\dagger} (0)  \ket{0}
  = \int_{\varepsilon_{min}}^\infty \frac{dp_0}{2\pi i} e^{-p_0\tau} {\rm Disc} D^{ij}(p_0).
\end{align}
${\cal O}^i$ and ${\cal O}^j$ are interpolation operators  with momenta ${\boldsymbol p}_i$ and ${\boldsymbol p}_j$, respectively (${\boldsymbol p}_i={\boldsymbol p}_j\equiv{\boldsymbol p}$), which are chosen depending on the channel of interest, $\varepsilon_{min}$ is the minimum energy and ${\rm Disc} D^{ij}(p_0) = D^{ij}(p_0+i\varepsilon) - D^{ij}(p_0-i\varepsilon)$.
As a function of $p_0$, $D^{ij}(p_0)$ has a branch cut starting from each multi-hadron threshold and is multi-valued.
It is known, however, that as a function of an appropriate variable $D^{ij}(p_0)$ becomes single-valued.
Such a procedure is called uniformization \cite{Newton:book}.
The uniformization variable, $u$,  for the single and two-channel systems can be given as follows.
\begin{itemize}
\item
single-channel with $\varepsilon = m+m'$, where $m$ and $m'$ are masses of hadrons in the channel:
\begin{align}
  u = k = \frac{1}{2}(p_0^2 - \boldsymbol p^2 - \varepsilon^2)^{1/2}.
\end{align}
\item
two-channel with $\varepsilon_1=m_1+m_1' < \varepsilon_2=m_2+m_2'$, where $m_1$ and $m_1'$ ($m_2$ and $m_2'$) are masses of hadrons in the channel 1 (2):
\begin{align}
  u = z = \frac{k_1+k_2}{\Delta} = \frac{1}{2} \frac{(p_0^2 - \boldsymbol p^2 - \varepsilon_1^2)^{1/2} + (p_0^2 - \boldsymbol p^2 - \varepsilon_2^2)^{1/2}}{\Delta},
\end{align}
where $\Delta = \frac{1}{2}\sqrt{\varepsilon_2^2-\varepsilon_1^2}$ \cite{NEWTON195829,Kato:1965iee}.
\end{itemize}
In terms of the uniformization variable, $u$, $D^{ij}(u)$ can be expressed as a sum of the pole terms, the Mittag-Leffler expansion \cite{arfken2013mathematical,whittaker1928course}, as
\begin{align}
  D^{ij} (u) &= \sum_n \left( \frac{r_n^{ij}}{u - u_n} - \frac{r_n^{ji*}}{u + u_n^*} \right).
\end{align}
Substituting Eq.(4) into Eq.(1) we obtain
\begin{align}
  C^{ij} (\tau) &= {\rm Disc} \sum_n \left( r_n^{ij} C (\tau,u_n) - r_n^{ji*} C (\tau,-u_n^*) \right),
\end{align}
where
\begin{align}
  C(\tau,u_n) = \int_{\varepsilon_{min}}^\infty \frac{dp_0}{2\pi i} e^{-p_0\tau}  \frac{1}{u(p_0) - u_n}.
\end{align}
It is worth while special attention that the imaginary-time correlation function is parametrized only by the pole positions and residues. 
Therefore, we can determine the pole positions and residues by fitting Eq.\,(5) together with Eq.\,(6) to the results of the lattice QCD simulations.
The pole position in the uniformization variable directly gives the complex energy of the unstable state and the Riemann sheet on which the unstable state is located and the residue is related to the information, such as the form factor (coupling constant) or the scattering amplitude. 
In the following we explain more details of the pole expansion of the imaginary-time correlation function taking specific examples of single- and two-channels.

First, we take the $\rho$ meson as an example of the unstable state in the $\pi\pi$ single-channel, which has been studied most extensively by lattice QCD simulations \cite{CP-PACS:2007wro,Feng:2010es,CS:2011vqf,Lang:2011mn,Dudek:2012xn,Wilson:2015dqa,Alexandrou:2017mpi,Andersen:2018mau}.
Following Ref.\,\cite{CP-PACS:2007wro} we consider two interpolating operators,
\begin{equation}
\begin{split}
  {\cal O}^\rho_\mu(\boldsymbol p,\tau) =& \rho_\mu(\boldsymbol p,\tau) \\
  {\cal O}^{\pi\pi}(\boldsymbol q_1,\boldsymbol q_2,\tau) =& \frac{1}{\sqrt{2}} \left( \pi^-({\boldsymbol q}_1,\tau) \pi^+({\boldsymbol q}_2,\tau) \right.
  \left. - \pi^+({\boldsymbol q_1},\tau) \pi^-({\boldsymbol q_2},\tau) \right),
\end{split}
\end{equation}
where $\rho_\mu(\boldsymbol p,\tau)$ and $\pi({\boldsymbol q},\tau)$ are interpolating operators for the $\rho$ and $\pi$ mesons with the spatial momentum $\boldsymbol p$ and $\boldsymbol q$, respectively.
Then, we construct a $2 \times 2$ matrix of the imaginary-time correlation function in the C.M.\,frame, $\boldsymbol p=0$,
\begin{align}
  \begin{pmatrix}
  {\cal C}^{\rho\rho}_{\mu\nu} (\tau) & {\cal C}^{\rho\pi\pi}_{\mu}(\tau) \\
  {\cal C}^{\pi\pi\rho}_\nu (\tau) & {\cal C}^{\pi\pi\pi\pi}(\tau) 
  \end{pmatrix}
&=
  \begin{pmatrix}
  \bra{0} {\cal O}^\rho_\mu(\boldsymbol 0,\tau)  {\cal O}^{\rho\dagger}_\nu (\boldsymbol 0,0)  \ket{0} & \bra{0} {\cal O}^{\rho}_\mu(\boldsymbol 0,\tau)  {\cal O}^{\pi\pi\dagger} (\boldsymbol q', -\boldsymbol q',0)  \ket{0} \\
  \bra{0} {\cal O}^{\pi\pi}(\boldsymbol q,-\boldsymbol q,\tau)  {\cal O}^{\rho\dagger}_\nu (\boldsymbol 0,0)  \ket{0} & \bra{0} {\cal O}^{\pi\pi}(\boldsymbol q,-\boldsymbol q,\tau)  {\cal O}^{\pi\pi\dagger} (\boldsymbol q', -\boldsymbol q',0)  \ket{0}
  \end{pmatrix}
\notag\\
&=
  \begin{pmatrix}
  \displaystyle{ \int_{2m_\pi}^\infty \frac{dp_0}{2\pi i} e^{-p_0\tau} {\rm Disc} {\cal D}^{\rho\rho}_{\mu\nu}(p_0)} & \displaystyle{ \int_{2m_\pi}^\infty \frac{dp_0}{2\pi i} e^{-p_0\tau} {\rm Disc} {\cal D}^{\rho\pi\pi}_\nu(p_0)} \\
  \displaystyle{ \int_{2m_\pi}^\infty \frac{dp_0}{2\pi i} e^{-p_0\tau} {\rm Disc} {\cal D}^{\pi\pi\rho}_\mu(p_0)} & \displaystyle{ \int_{2m_\pi}^\infty \frac{dp_0}{2\pi i} e^{-p_0\tau} {\rm Disc} {\cal D}^{\pi\pi\pi\pi}(p_0)} 
  \end{pmatrix}
\end{align}
From ${\cal D}^{\rho\rho}_{\mu\nu}$, ${\cal D}^{\pi\pi\rho}_\mu$, ${\cal D}^{\rho\pi\pi}_\nu$ and ${\cal D}^{\pi\pi\pi\pi}$, we define $D^{\rho\rho}$, $D^{\pi\pi\rho}$, $D^{\rho\pi\pi}$ and $D^{\pi\pi\pi\pi}$
for convenience by
  ${\cal D}^{\rho\rho}_{ij} =  \delta_{ij} D^{\rho\rho}$, ${\cal D}^{\pi\pi\rho}_i = 2 q_i D^{\pi\pi\rho}$, ${\cal D}^{\rho\pi\pi}_i = 2 q'_i D^{\rho\pi\pi}$ and
  ${\cal D}^{\pi\pi\pi\pi} = 4 \boldsymbol q \cdot \boldsymbol q' D^{\pi\pi\pi\pi}$ (and accordingly $C^{\rho\rho}$, $C^{\pi\pi\rho}$, $C^{\rho\pi\pi}$ and $C^{\pi\pi\pi\pi}$).
$D^{\rho\rho}$, $D^{\pi\pi\rho}$, $D^{\rho\pi\pi}$ and $D^{\pi\pi\pi\pi}$ are diagrammatically represented as
\begin{align}
  \begin{pmatrix}
    D^{\rho\rho}(p_0) & D^{\rho\pi\pi}(p_0) \\
    D^{\pi\pi\rho}(p_0) & D^{\pi\pi\pi\pi}(p_0)
  \end{pmatrix}
  = \left(
        \raisebox{-15mm}{\includegraphics[width=0.5\linewidth]{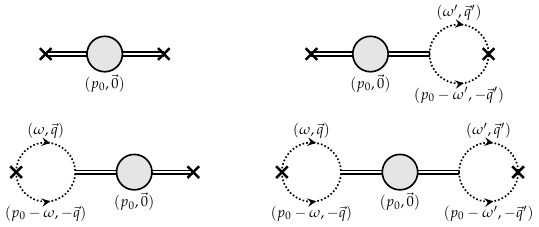}}
  \right)
\end{align}
As a function of the uniformization variable, $k$, defined by Eq.\,(2),
the pole expansion of $D^{\rho\rho}$, $D^{\pi\pi\rho}$, $D^{\rho\pi\pi}$ and $D^{\pi\pi\pi\pi}$ is given as
\begin{equation}
\begin{split}
  D^{\rho\rho} &= \sum_\rho \left( \frac{r_\rho^{\rho\rho}}{k-k_\rho} -  \frac{r_\rho^{\rho\rho *}}{k+k_\rho^*} \right) \\
  D^{\pi\pi\rho} &= \sum_{\pi\pi} \left( \frac{r_{\pi\pi}^{\pi\pi\rho}}{k-k_{\pi\pi}} - \frac{r_{\pi\pi}^{\rho\pi\pi*}}{k+k_{\pi\pi}^*} \right) + \sum_\rho \left( \frac{r_\rho^{\pi\pi\rho*}}{k-k_\rho} -  \frac{r_\rho^{\rho\pi\pi *}}{k+k_\rho^*} \right) \\
  D^{\rho\pi\pi} &= \sum_{\pi\pi'} \left( \frac{r_{\pi\pi'}^{\rho\pi\pi}}{k-k_{\pi\pi'}} - \frac{r_{\pi\pi'}^{\pi\pi\rho*}}{k+k_{\pi\pi'}^*} \right) + \sum_\rho \left( \frac{r_\rho^{\rho\pi\pi}}{k-k_\rho} -  \frac{r_\rho^{\pi\pi\rho *}}{k+k_\rho^*} \right) \\
  D^{\pi\pi\pi\pi} &= \sum_{\pi\pi} \left( \frac{r_{\pi\pi}^{\pi\pi\pi\pi}}{k-k_{\pi\pi}} - \frac{r_{\pi\pi}^{\pi\pi\pi\pi*}}{k+k_{\pi\pi}^*} \right) + \sum_{\pi\pi'} \left( \frac{r_{\pi\pi'}^{\pi\pi\pi\pi}}{k-k_{\pi\pi'}} - \frac{r_{\pi\pi'}^{\pi\pi\pi\pi*}}{k+k_{\pi\pi'}^*} \right) + \sum_\rho \left( \frac{r_\rho^{\pi\pi\pi\pi}}{k-k_\rho} -  \frac{r_\rho^{\pi\pi\pi\pi *}}{k+k_\rho^*} \right) 
\end{split}
\end{equation}
The poles at $k = k_\rho$ and $-k_\rho^*$, are dynamical and due to the propagation of the $\rho$ and higher unstable states, \raisebox{-1.5mm}{\includegraphics[width=0.1\linewidth]{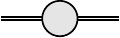}}, which are independent of the momentum of the interpolating field of the pions, $\boldsymbol q$ ($\boldsymbol q'$). The lowest pole position gives the mass and width of $\rho$ as
$m_\rho - i\Gamma_\rho/2 = \left(4m_\pi^2 + 4k_\rho^2\right)^{1/2}$.
The poles at $k = k_{\pi\pi}$ and $-k_{\pi\pi}^*$ ($k = k_{\pi\pi'}$ and $-k_{\pi\pi'}^*$) are kinematical, due to the propagation of noninteracting $\pi\pi$ and higher unstable states, \raisebox{-5.5mm}{\includegraphics[width=0.05\linewidth]{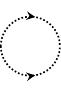}}, which are dependent on $\boldsymbol q$ ($\boldsymbol q'$), $k_{\pi\pi} = \left|\boldsymbol q \right | - i \epsilon$ ($k_{\pi\pi'} = \left|\boldsymbol q' \right | - i \epsilon$) for the lowest $\pi\pi$ pole.
On the other hand, the residue of the lowest $\pi\pi$ pole gives the form factor and that of the lowest $\rho$ pole includes the $\pi\pi$ scattering volume for $\boldsymbol q = 0$.
We note here that $P$ and $CP$ symmetry results in $D^{\pi\pi\rho} = D^{\rho\pi\pi}$, which means $r^{\pi\pi\rho}=r^{\rho\pi\pi}$, and $C^{\pi\pi\rho} = C^{\rho\pi\pi}$ (real) for $\boldsymbol q=\boldsymbol q'$ in the present definition.

In Ref.\,\cite{Maiani:1990ca} Maiani and Testa studied the three-point imaginary-time correlation function and derived a theorem relating their asymptotic behavior and the scattering length (volume).
In our notation the three-point imaginary-time correlation function is
\begin{align}
  \bra{0} \pi(\boldsymbol q, \tau_1) \pi(-\boldsymbol q, \tau_2) J_i (0)  \ket{0}
  &= 2q_i C^{\pi\pi\rho} (\tau_1,\tau_2).
\end{align}
$J_\mu$ is the vector current, which plays a role of the interpolating operator for the $\rho$ meson, $\rho_\mu$ of Eq.\,(7) up to normalization.
They showed that for $\boldsymbol q = 0$, $\tau_1 \rightarrow \infty$ and $0 \ll \tau_2 \ll \tau_1$,
\begin{align}
  C^{\pi\pi\rho} (\tau_1,\tau_2)
  &\approx \frac{Z_\pi}{2m_\pi^2}  e^{-m_\pi\tau_1} e^{-m_\pi\tau_2} f(4m_\pi^2) \left( 1 - \frac{1}{\sqrt{2}}\frac{v m_\pi}{6\tau_2}  \sqrt{\frac{m_\pi}{4\pi\tau_2}} + \cdots \right),
\end{align}
where $Z_\pi$ is the pion-field renormalization, $v$ is the $\pi\pi$ scattering volume, related to the phase shift as $\delta = \frac{1}{3}vq^3 + \cdots$, and $f$ is the vector form factor
\begin{align}
 2 E_{\boldsymbol q}  \bra {\pi^+(\boldsymbol q) \pi^-(-\boldsymbol q) \,  \text{out}} J_i(0) \ket{0 } = q_i f(4 E_{\boldsymbol q}^2) ,
\end{align}
while for $\boldsymbol q \neq 0$, the coefficient of the negative power of $\tau_2$ is  proportional to an off-shell amplitude with no direct meaning in terms of observable quantities.

For $\boldsymbol q = 0$, the pole expansion of $D^{\pi\pi\rho}$ is
\begin{align}
  D^{\pi\pi\rho} &= \frac{s_{\pi\pi}^{\pi\pi\rho}}{k^2} + {\sum_{\pi\pi}}' \left( \frac{r_{\pi\pi}^{\pi\pi\rho}}{k-k_{\pi\pi}} - \frac{r_{\pi\pi}^{\pi\pi\rho*}}{k+k_{\pi\pi}^*} \right)  + \sum_\rho \left( \frac{r_\rho^{\pi\pi\rho}}{k-k_\rho} -  \frac{r_\rho^{\pi\pi\rho *}}{k+k_\rho^*} \right).
\end{align}
The first term of $D^{\pi\pi\rho}$ in Eq.\,(10) is explicitly separated into the lowest $\pi\pi$ pole contribution, the first term, and the rest, the second term in Eq.\,(14).
Just above the $\pi\pi$ threshold, ${\rm Im}D^{\pi\pi\rho}$ behaves as
\begin{align}
  {\rm Im}D^{\pi\pi\rho} &=  -s_{\pi\pi}^{\pi\pi\rho} \pi \delta(k^2) - 2{\rm Im} \left( {\sum_{\pi\pi}}'\frac{r_{\pi\pi}^{\pi\pi\rho}}{k_{\pi\pi}^2} + \sum_{\rho}\frac{r_\rho^{\pi\pi\rho}}{k_\rho^2} \right) k + \cdots. 
\end{align}
Then, the long-time behavior of $C^{\pi\pi\rho}$ is
\begin{align}
  C^{\pi\pi\rho} (\tau)
  &\approx -\frac{s_{\pi\pi}^{\pi\pi\rho}}{4 m_\pi} e^{-2m_\pi\tau} - {\rm Im} \left( {\sum_{\pi\pi}}'\frac{r_{\pi\pi}^{\pi\pi\rho}}{k_{\pi\pi}^2} + \sum_{\rho}\frac{r_\rho^{\pi\pi\rho}}{k_\rho^2} \right)  e^{-2m_\pi\tau} \tau^{-3/2} \sqrt{\frac{m_\pi}{4\pi}} \nonumber\\
  &\approx -\frac{s_{\pi\pi}^{\pi\pi\rho}}{4 m_\pi} e^{-2m_\pi\tau} \left( 1 -\frac{2}{3}  \frac{v  m_\pi}{\tau} \sqrt{\frac{m_\pi}{4\pi\tau}} \right),
\end{align}
where the scattering volume, $v$, is related to the residues of the poles together with the pole position as
\begin{align}
  & v = \lim_{k \to 0} \frac{3}{k^3} \frac{{\rm Im}\left[k^2  D^{\pi\pi\rho}\right]}{ {\rm Re}\left[k^2 D^{\pi\pi\rho}\right]} = - \frac{6}{s_{\pi\pi}^{\pi\pi\rho}}\, {\rm Im} \left( {\sum_{\pi\pi}}'\frac{r_{\pi\pi}^{\pi\pi\rho}}{k_{\pi\pi}^2} + \sum_{\rho}\frac{r_\rho^{\pi\pi\rho}}{k_\rho^2} \right).
\end{align}
Eq.\,(16) should be compared with Eq.\,(12).
In the former $\tau_2 = \tau_1=\tau$ while in the latter $\tau_2 \ll \tau_1$, though both $\tau_1$ and $\tau_2$ are taken to be large in both cases.

Now, we demonstrate the pole expansion of the imaginary-time correlation function adopting the vector dominance model \cite {Sakurai:1960ju} with the interaction Lagrangian,
\begin{align}
  {\cal L} = i g \rho^\mu \left( \pi^+ \partial_\mu \pi^- - \pi^- \partial_\mu \pi^+  \right).
\end{align}
We follow Ref.\,\cite{Klingl:1996by} for the details of the calculation with the parameters,  $m_\pi=0.140$ GeV, $m_\rho=0.760$ GeV, $\Gamma=0.160$ GeV and $g=6.05$.
We consider two cases of $|\boldsymbol q|=0$ and $1$ GeV.

The noninteracting $\pi\pi$ propagation is trivially given by a pair of poles in the present demonstration
and we found only a pair of $\rho$ poles in the region of interest.
Table \ref{tab:rho_poles} shows the positions and the residues of the poles in the uniformization variable, $k$, together with the complex pole energy
and Fig.\,1 shows the pole positions in the complex $k$ plane.

\begin{table}[h!]
\caption{List of poles of $D^{\pi\pi\rho}$ in the vicinity of interest for cases $|\boldsymbol q|=0$ (above) and $1$ GeV (below).
Pole positions are given in terms of the uniformization variable $k$ and energy $E$. 
Note that residues $r^{\pi\pi\rho}$ ($s^{\pi\pi\rho}$) are the values on the $k$ plane.
}\label{tab:rho_poles}
\vspace{10pt}\par
\begin{tabular}{c|ccc}\hline\hline
pole & $k$ [GeV] & $E$ [GeV] & $r^{\pi\pi\rho} [\text{GeV}^{-4}] (s^{\pi\pi\rho} [\text{GeV}^{-3}])$ \\\hline
$\rho$ & $\pm 0.354 - 0.039 i$ & $0.760 \mp 0.073 i$ & $\pm 28.93 + 2.27 i $ \\
$\pi\pi$ & $0$ & $0.279$ & $-19.18$\\ \hline\hline
\end{tabular}
\vspace{15pt}\par
\begin{tabular}{c|ccc}\hline\hline
pole & $k$ [GeV] & $E$ [GeV] & $r^{\pi\pi\rho} (s^{\pi\pi\rho}) [\text{GeV}^{-4}] (s^{\pi\pi\rho} [\text{GeV}^{-3}])$ \\ \hline
$\rho$ & $\pm 0.354 - 0.039 i$ & $0.760 \mp 0.073 i$ & $\mp 0.571 + 0.101 i$ \\
$\pi\pi$ & $\pm 1$ & $2.019$ & $0.221 - 0.068 i$\\ \hline\hline
\end{tabular}
\end{table}
\begin{figure}[h!]
      \centering
      \includegraphics[width=\linewidth]{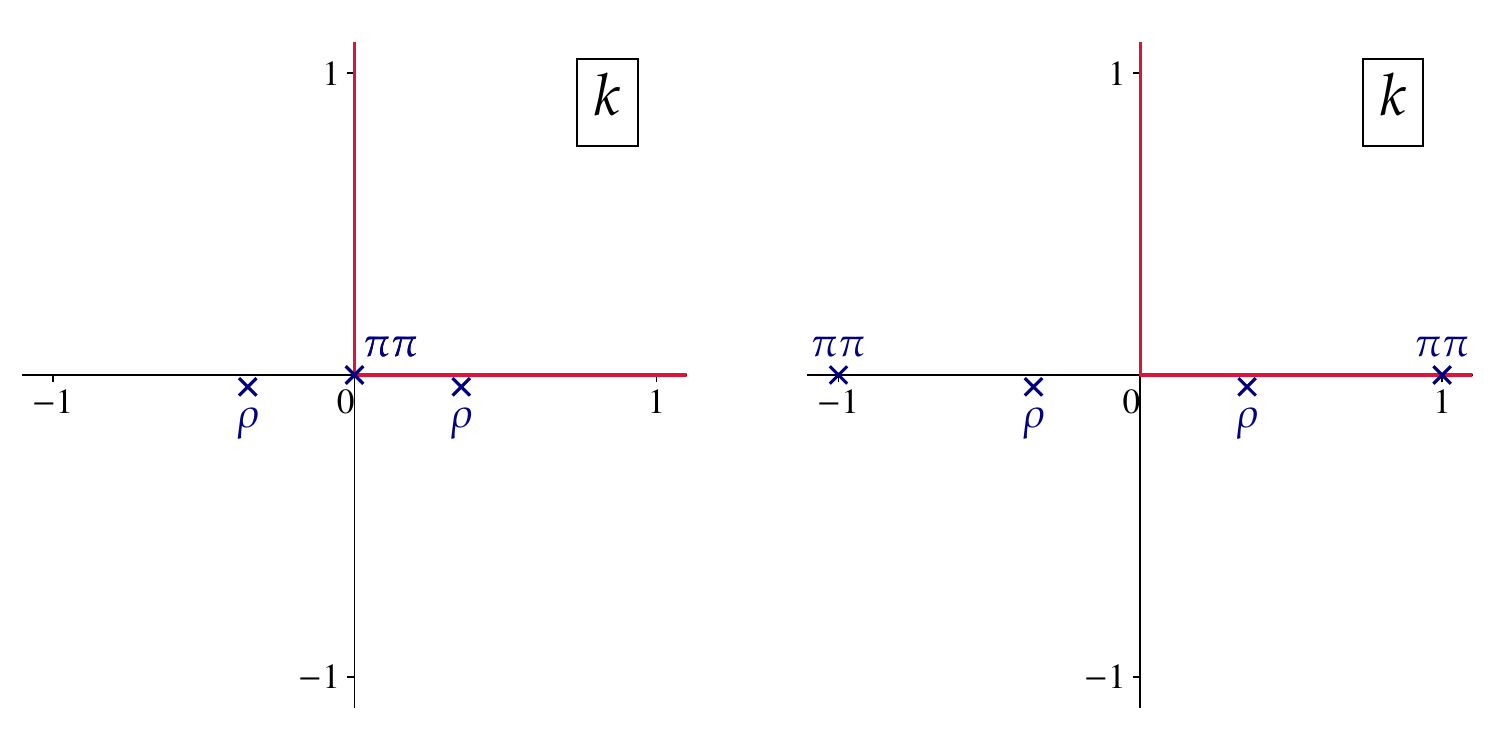}
      \caption{Pole positions of $D^{\pi\pi\rho}$ in the $k$ plane for cases $|\boldsymbol q|=0$ (left) and $1$ GeV (right).
      The labels correspond to the poles listed in Table \ref{tab:rho_poles}.
      The red line shows the physical region.}\label{fig:k-plane}
\end{figure}

We show ${\rm Re}D^{\pi\pi\rho}$, ${\rm Im}D^{\pi\pi\rho}$ and $C^{\pi\pi\rho}$ in Figs.\,2, 3 and 4, respectively.
The direct calculation and the sum of contributions of the $\rho$ pole and the $\pi\pi$ pole almost perfectly coincide not only for $D^{\pi\pi\rho}$ but also for $C^{\pi\pi\rho}$, for $|\boldsymbol q| = 0$ and $1$ GeV.
For $|\boldsymbol q| = 0$, the energy of the $\pi\pi$ pole is much lower than that of the $\rho$ pole, $E_{\pi\pi} \ll m_\rho$, and $C^{\pi\pi\rho}$ is dominated by the $\pi\pi$ pole contribution.
While, for $|\boldsymbol q|=1$ GeV, the energy of the $\rho$ pole is much lower than that of the $\pi\pi$ pole, $m_\rho \ll E_{\pi\pi}$, and $C^{\pi\pi\rho}$ is dominated by the $\rho$ pole contribution.
Therefore, the pole position of the $\rho$ meson would be efficiently obtained from $C^{\pi\pi\rho}$ for  large values of $q$ ($m_\rho \ll E_{\pi\pi}$) though in principle it can be obtained for any value of $q$ if the $\pi\pi$ pole and the $\rho$ pole contributions can be separated in $C^{\pi\pi\rho}$.

\begin{figure}[ht!]
  \centering
  \begin{tabular}{cc}
    \begin{minipage}{0.5\hsize}
      \centering
      \includegraphics[width=\linewidth]{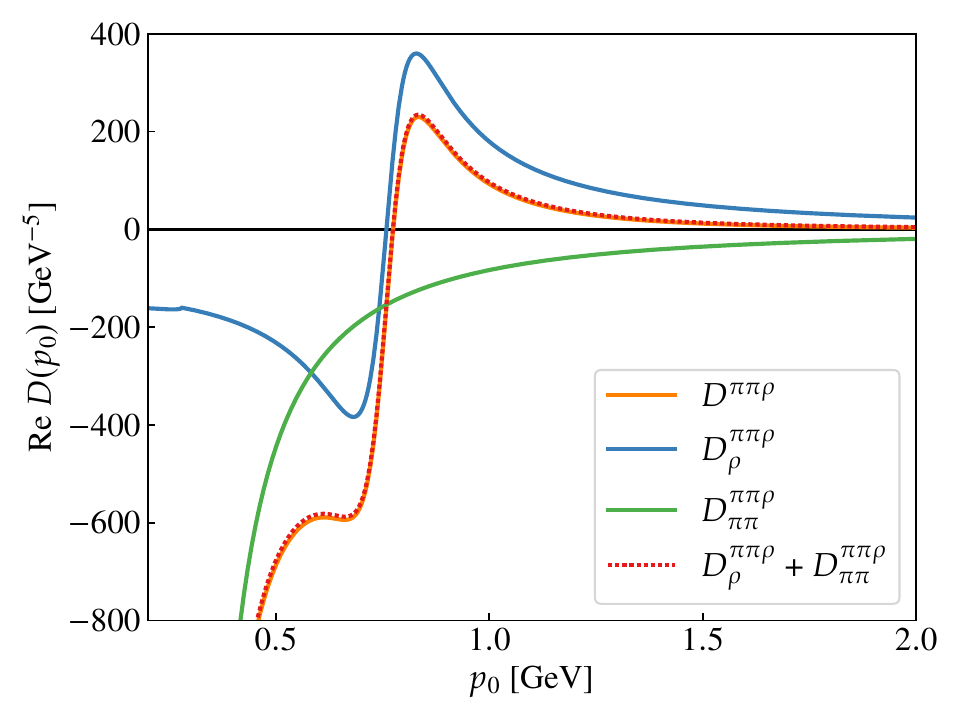}
    \end{minipage}
    \begin{minipage}{0.5\hsize}
      \centering
      \includegraphics[width=\linewidth]{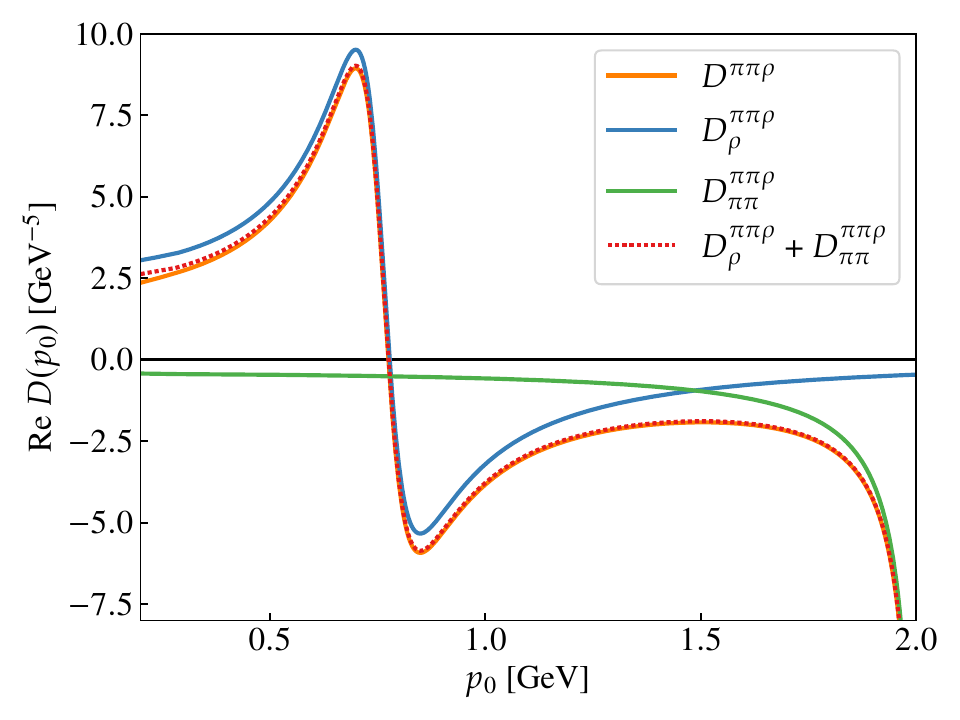}
    \end{minipage}
  \end{tabular}
\caption{$p_0$ dependence of the real part of $D^{\pi\pi\rho}$ and its pole contributions for cases $|\boldsymbol q|=0$ (left) and $1$ GeV (right).
The exact (total) $D^{\pi\pi\rho}$ is shown by the orange line.
The blue, green and red-dotted lines correspond to the contribution from the $\rho$ pole, the contribution from the $\pi\pi$ pole and the sum of the two contributions, respectively.
}\label{fig:re_rho}
\end{figure}
\begin{figure}[h!]
  \centering
  \begin{tabular}{cc}
    \begin{minipage}{0.5\hsize}
      \centering
      \includegraphics[width=\linewidth]{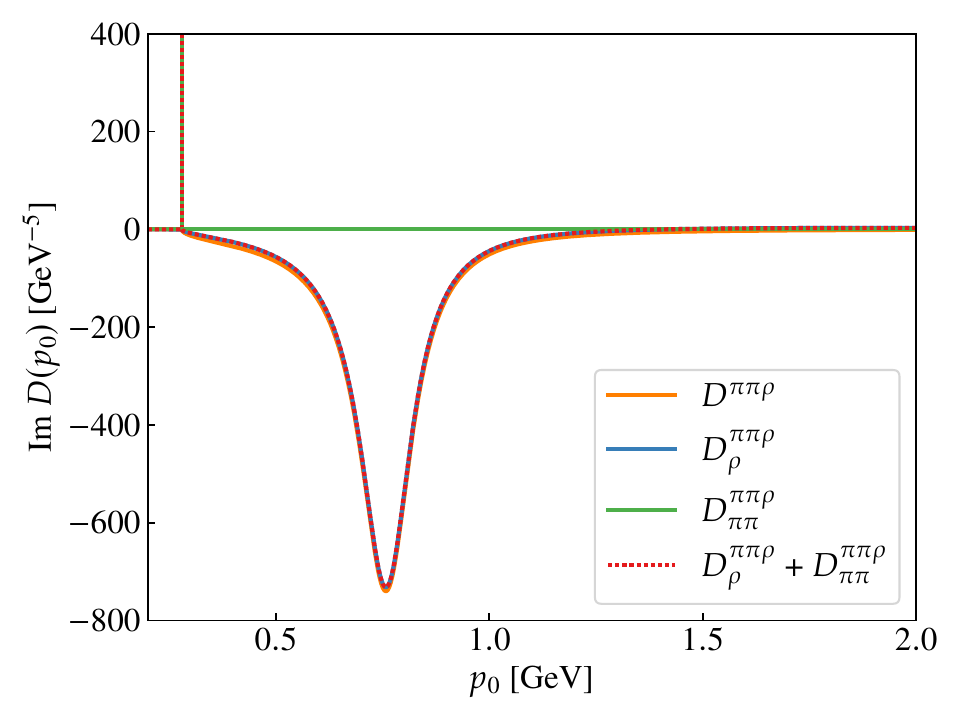}
    \end{minipage}
    \begin{minipage}{0.5\hsize}
      \centering
      \includegraphics[width=\linewidth]{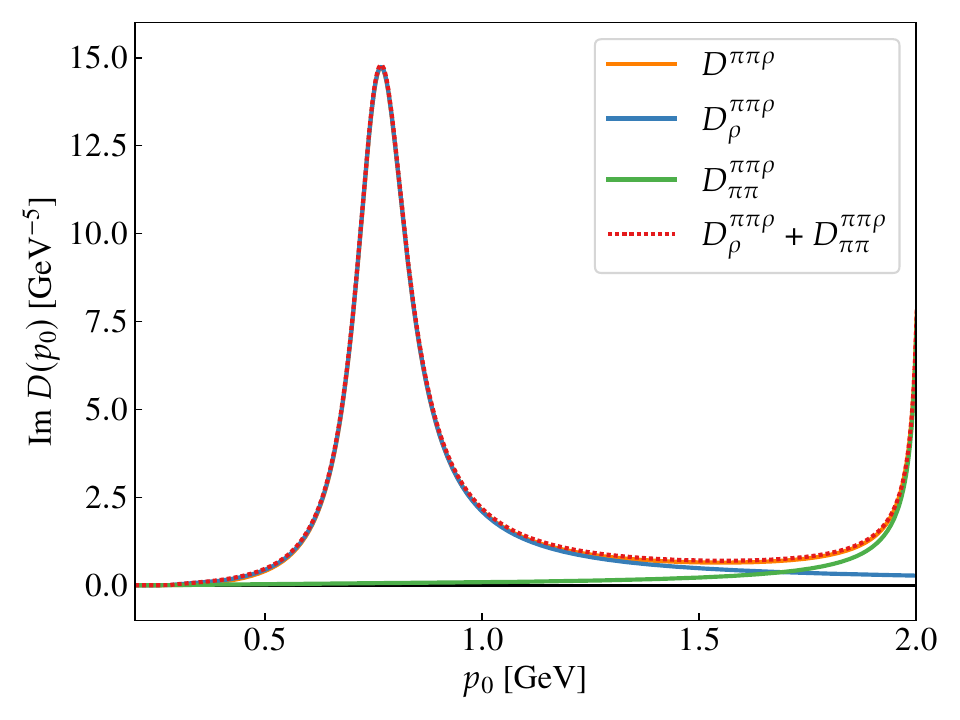}
    \end{minipage}
  \end{tabular}
\caption{$p_0$ dependence of the imaginary part of $D^{\pi\pi\rho}$ and its pole contributions for cases $|\boldsymbol q|=0$ (left) and $1$ GeV (right).
Each line has the same correspondence as that of Fig.\,\ref{fig:re_rho}.
}\label{fig:im_rho}
\end{figure}
\begin{figure}[htpb]
  \centering
  \begin{tabular}{cc}
    \begin{minipage}{0.5\hsize}
      \centering
      \includegraphics[width=\linewidth]{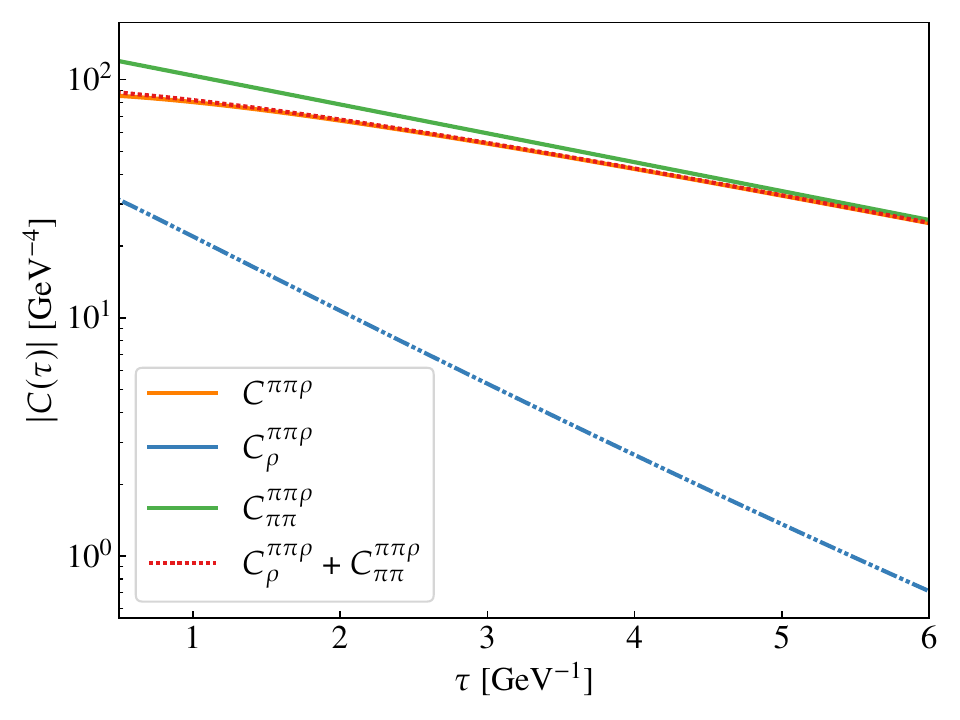}
    \end{minipage}
    \begin{minipage}{0.5\hsize}
      \centering
      \includegraphics[width=\linewidth]{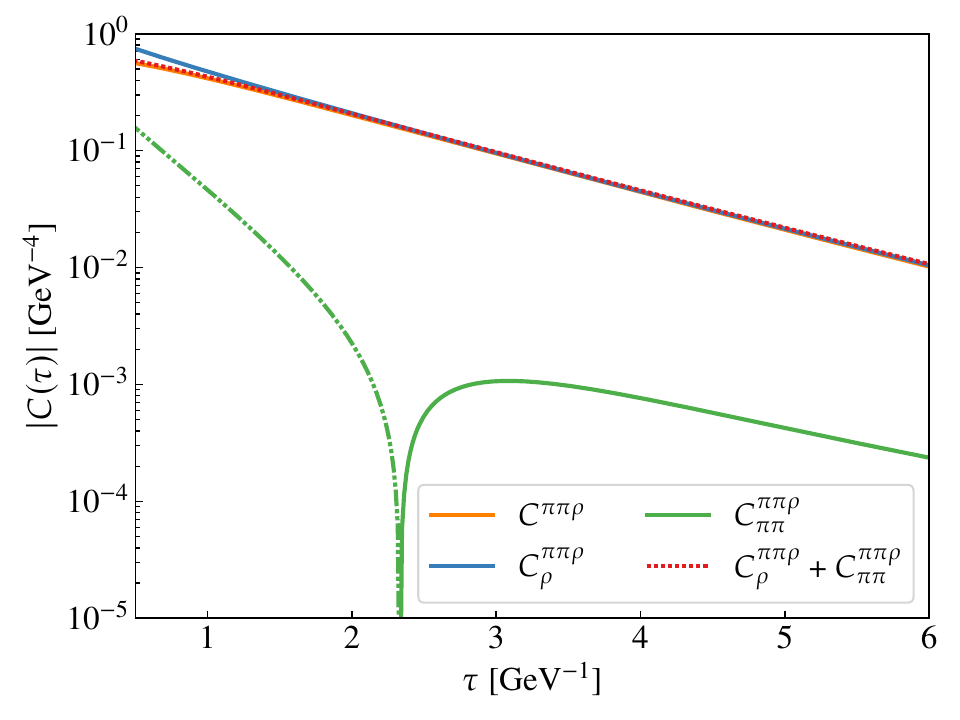}
    \end{minipage}
  \end{tabular}
\caption{$\tau$ dependence of the absolute value of $C^{\pi\pi\rho}$ and its pole contributions for cases $|\boldsymbol q|=0$ (left) and $1$ GeV (right) in log scale. The dot-dashed lines represent contributions with negative values.}
\label{fig:c_corr_log}
\end{figure}

Now, we would like to discuss how the mass and the width affect $C^{\pi\pi\rho}$. 
Fig.\,5 shows how the contribution of the $\rho$ pole is affected if the mass, $m_\rho$, and the width, $\Gamma_\rho$, change $\pm 20 \%$ for $|\boldsymbol q|=1$ GeV. One sees that the change of the mass is reflected as the change of the shape, mainly as the slope, while that of the width is reflected in the magnitude of the contribution but hardly in the change of the shape.
In fact the residues, $r_\rho^{\rho\rho}$, $r_\rho^{\rho\pi\pi}$, $r_\rho^{\pi\pi\rho}$ and $r_\rho^{\pi\pi\pi\pi}$ respectively depend on the coupling constant, $g$, as ${\cal O}(g^0)$, ${\cal O}(g)$, ${\cal O}(g)$ and ${\cal O}(g^2)$ and therefore on the decay width $\Gamma$ as ${\cal O}(\Gamma^0)$, ${\cal O}(\Gamma^{1/2})$,  ${\cal O}(\Gamma^{1/2})$ and ${\cal O}(\Gamma)$.

\begin{figure}[htpb]
  \centering
  \begin{tabular}{cc}
    \begin{minipage}{0.5\hsize}
      \centering
      \includegraphics[width=\linewidth]{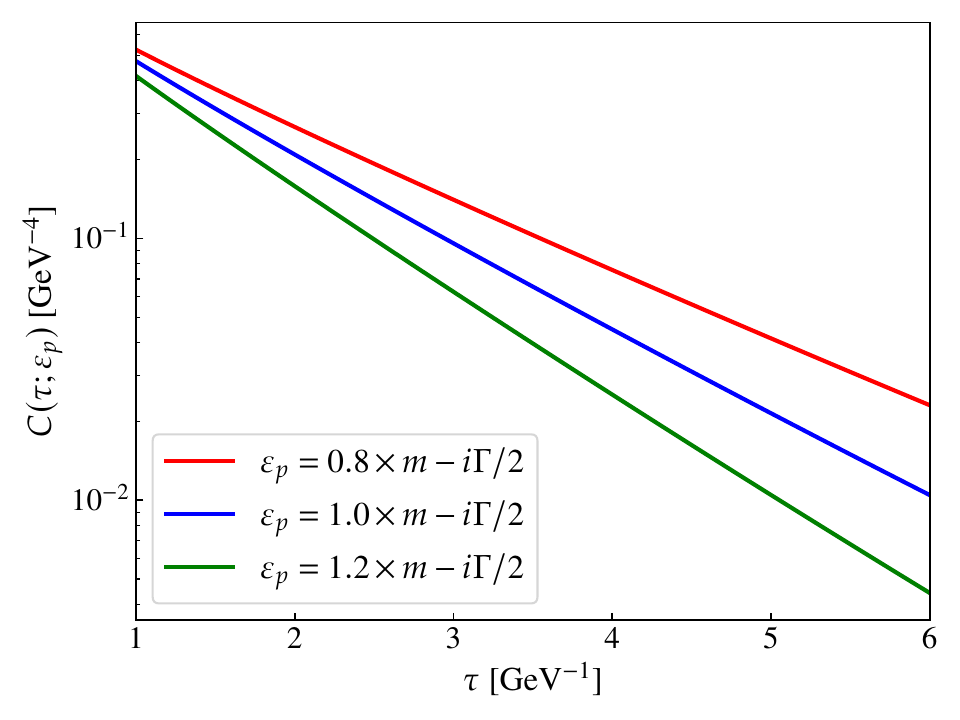}
    \end{minipage}
    \begin{minipage}{0.5\hsize}
      \centering
      \includegraphics[width=\linewidth]{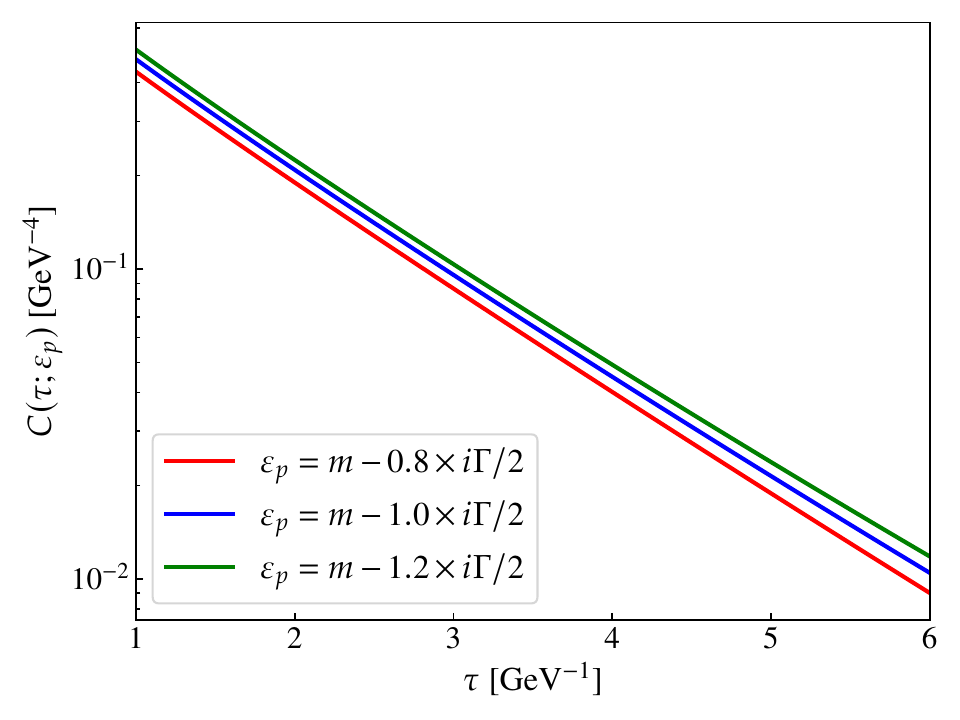}
    \end{minipage}
  \end{tabular}
\caption{Dependence of pole contribution $C^{\pi\pi\rho}_\rho$ on mass $M$ (left) and width $\Gamma$ (right) with $|\boldsymbol q|=1$ GeV.}
\label{fig:dept-m_gamma}
\end{figure}

Secondly, we take $\Lambda(1405)$ as an example of the unstable state in the $\pi\Sigma$-${\bar K}N$ coupled-channels, which has recently been studied by lattice QCD simulations \cite{BaryonScatteringBaSc:2023zvt,BaryonScatteringBaSc:2023ori}.
We consider two interpolating fields
\begin{equation}
\begin{split}
  {\cal O}^{\bar KN}(t) &= \frac{1}{\sqrt{2}} \left( K^-(\boldsymbol q,\tau) p(-\boldsymbol q,\tau) - \bar{K}^0(\boldsymbol q,\tau) n(-\boldsymbol q,\tau) \right), \\
  {\cal O}^{\pi\Sigma}(t) &= \frac{1}{\sqrt{3}} \left( \pi^-(\boldsymbol q',\tau) \Sigma^+(-\boldsymbol q',\tau) + \pi^0(\boldsymbol q',\tau) \Sigma^0(-\boldsymbol q',\tau) + \pi^+(\boldsymbol q',\tau) \Sigma^-(-\boldsymbol q',\tau) \right),
\end{split}
\end{equation}
where $h(\boldsymbol q,\tau)$ represents the interpolating operator for the hadron, $h$, on the right-hand side.
\begin{align}
  \begin{pmatrix}
  C^{\bar{K}N\bar{K}N} (\tau) & C^{\bar{K}N\pi\Sigma}(\tau) \\
  C^{\pi\Sigma\bar{K}N} (\tau) & C^{\pi\Sigma\pi\Sigma}(\tau) 
  \end{pmatrix}
&=
  \begin{pmatrix}
  \bra{0} {\cal O}^{\bar{K}N}(\tau)  {\cal O}^{\bar{K}N\dagger} (0)  \ket{0} & \bra{0} {\cal O}^{\bar{K}N}(\tau)  {\cal O}^{\pi\Sigma\dagger} (0)  \ket{0} \\
  \bra{0} {\cal O}^{\pi\Sigma}(\tau)  {\cal O}^{\bar{K}N\dagger} (0)  \ket{0} & \bra{0} {\cal O}^{\pi\Sigma}(\tau)  {\cal O}^{\pi\Sigma\dagger} (0)  \ket{0}
  \end{pmatrix}
\nonumber\\
&=
  \begin{pmatrix}
  \displaystyle{ \int_0^\infty \frac{dp_0}{2\pi i} e^{-p_0\tau} {\rm Disc} D^{\bar{K}N\bar{K}N}(p_0)} & \displaystyle{ \int_0^\infty \frac{dp_0}{2\pi i} e^{-p_0\tau} {\rm Disc} D^{\bar{K}N\pi\Sigma}(p_0)} \\
  \displaystyle{ \int_0^\infty \frac{dp_0}{2\pi i} e^{-p_0\tau} {\rm Disc} D^{\pi\Sigma\bar{K}N}(p_0)} & \displaystyle{ \int_0^\infty \frac{dp_0}{2\pi i} e^{-p_0\tau} {\rm Disc} D^{\pi\Sigma\pi\Sigma}(p_0)} 
  \end{pmatrix}.
\end{align}
$D^{\bar{K}N\bar{K}N}$, $D^{\bar{K}N\pi\Sigma}$, $D^{\pi\Sigma\bar{K}N}$ and $D^{\pi\Sigma\pi\Sigma}$ are diagrammatically represented as
\begin{align}
  \begin{pmatrix}
    D^{\bar{K}N\bar{K}N}(p_0) & D^{\bar{K}N\pi\Sigma}(p_0) \\
    D^{\pi\Sigma\bar{K}N}(p_0) & D^{\pi\Sigma\pi\Sigma}(p_0)
  \end{pmatrix}
  = \left(
        \raisebox{-15mm}{\includegraphics[width=0.5\linewidth]{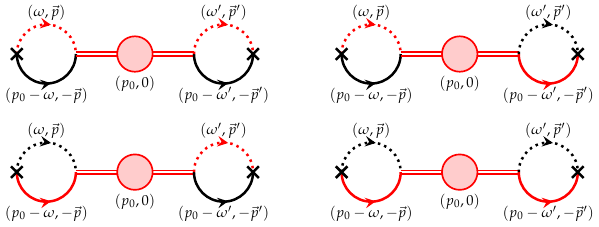}}
  \right).
\end{align}
Here, we consider $D^{\bar{K}N\pi\Sigma}$ only.
As a function of the uniformization variable, $z$, defined by Eq.\,(3)
the pole expansion of $D^{\bar{K}N\pi\Sigma}$ is
\begin{align}
  D^{\bar{K}N\pi\Sigma} &= \sum_{\bar{K}N}\left( \frac{r_{\bar{K}N}^{\bar{K}N\pi\Sigma}}{z-z_{\bar{K}N}} - \frac{r_{\bar{K}N}^{\bar{K}N\pi\Sigma*}}{z+z_{\bar{K}N}^*} \right) + \sum_{\pi\Sigma} \left( \frac{r_{\pi\Sigma}^{\bar{K}N\pi\Sigma}}{z-z_{\pi\Sigma}} -  \frac{r_{\pi\Sigma}^{\bar{K}N\pi\Sigma *}}{z+z_{\pi\Sigma}^*} \right) + \sum_\Lambda \left( \frac{r_\Lambda^{\bar{K}N\pi\Sigma}}{z-z_\Lambda} -  \frac{r_\Lambda^{\bar{K}N\pi\Sigma *}}{z+z_\Lambda^*} \right).
\end{align}
Similarly to the single-channel case, the poles at $z=z_\Lambda$ and $-z_\Lambda^*$ are due to the propagation of the $\Lambda(1405)$ and higher unstable states, \raisebox{-1.5mm}{\includegraphics[width=0.1\linewidth]{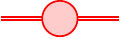}}, while poles at $z=z_{\bar{K}N}$ and $-z_{\bar{K}N}^*$ ($z=z_{\pi\Sigma}$ and $-z_{\pi\Sigma}^*$) are due to the propagation of noninteracting $\bar{K}N$, \raisebox{-5.5mm}{\includegraphics[width=0.05\linewidth]{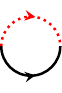}} ($\pi\Sigma${,  \raisebox{-5.5mm}{\includegraphics[width=0.05\linewidth]{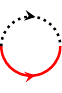}}).
\par
We demonstrate the pole expansion of the imaginary-time correlation function adopting the chiral unitary model \cite{Kaiser:1996js,Oset:1997it,Hyodo:2011ur,Morimatsu:2019wvk} with the interaction Lagrangian,
\begin{align}
    \mathcal{L}_{TW}=i\frac{C_{ji}}{4f^2}\overline{\psi}_j\phi_j^\dagger\overleftrightarrow{\cancel{\partial}}\phi_i\psi_i,
\end{align}
where $\phi=(\pi,\bar K)^T$ and $\psi=(\Sigma,N)^T$, respectively, $f=0.104$ GeV and
\begin{align}
C=
\begin{pmatrix}
3 &-\sqrt{\frac{3}{2}} \\
-\sqrt{\frac{3}{2}} & 4\\
\end{pmatrix}
.
\end{align}
We refer to Ref.\,\cite{Yamada:2020rpd} for the details of the calculation.
$|\boldsymbol q|$  and $|\boldsymbol q'|$ are taken to be 1 GeV.

As in the single-channel case, the noninteracting $\bar{K}N$ or $\pi\Sigma$ propagation is trivially given by a pair of poles
while we found three pairs of $\Lambda$ poles in the region of interest.
Table 2 shows the positions and the residues of the poles in the uniformization variable, $z$, together with the complex pole energy
and Fig.\,6 shows the pole positions in the $z$ plane.

\begin{table}[h!]
\caption{List of poles of $D^{{\bar K}N\pi\Sigma}$ in the vicinity of interest for $|\boldsymbol q|=1$ GeV and $|\boldsymbol q'|=1$ GeV.
Pole positions are given in terms of the uniformization variable, $z$, and energy, $E$. 
Note that residues, $r^{{\bar K}N\pi\Sigma}$, are the values on the $z$ plane.}\label{tab:lambda1405_poles}
\vspace{10pt}\par
\begin{tabular}{c|ccc}\hline\hline
pole & $z$ & $E$ [GeV] & $r^{\bar{K}N\pi\Sigma}$ $[\text{GeV}^{-5}]$\\ \hline
$\Lambda_1$ & $\pm 0.760 + 0.154 i $ & $1.435 \mp 0.010 i $ & $\pm 43.40 - 3.14 i$ \\
$\Lambda_2$ & $\pm 0.413 + 0.395 i $ & $1.386 \mp 0.072 i $ & $\mp 10.27 - 24.09  i$ \\
$\Lambda_3$ & $\pm 1.72 - 1.49 i $ & $1.397 \mp 0.131 i $ & $\mp18.11 + 17.45 i$ \\
$\pi\Sigma$ & $\pm 8.03$ & $2.57 $ & $\pm 9.35 - 4.34 i$ \\
$\bar K N$ & $\pm 7.68$ & $2.49$ & $\mp 11.64 + 5.16 i$\\ \hline\hline
\end{tabular}
\end{table}
\begin{figure}[h!]
  \centering
  \includegraphics[width=0.5\linewidth]{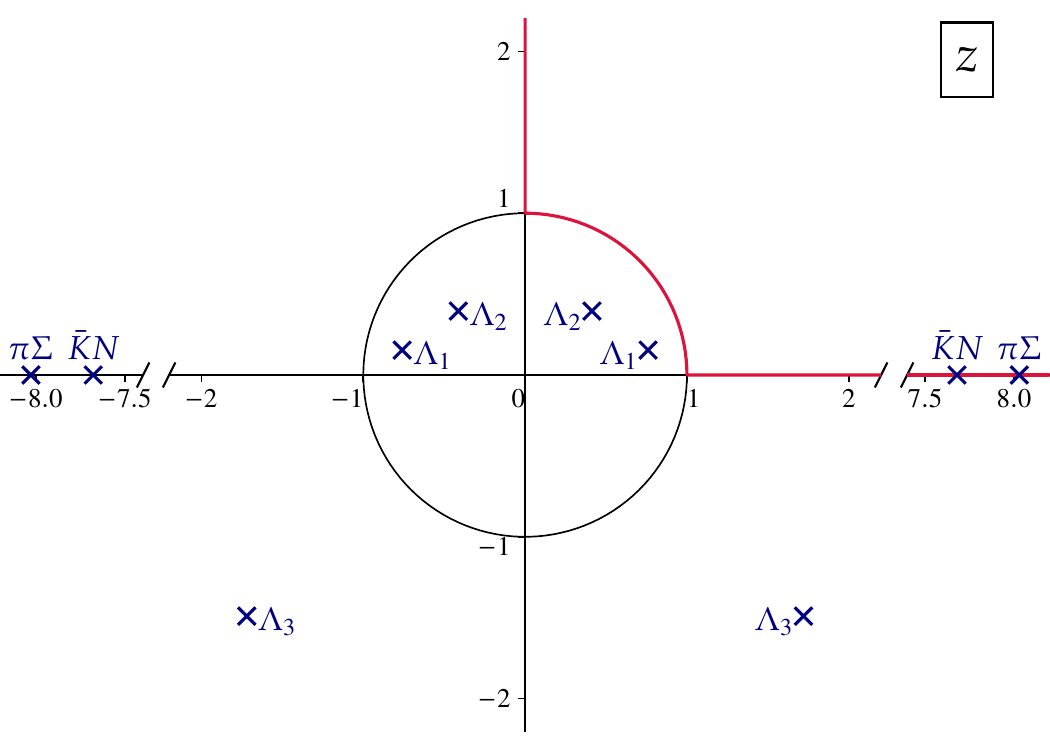}
  \caption{Pole positions of $D^{{\bar K}N\pi\Sigma}$ in the $z$ plane. 
  The labels correspond to the poles listed in Table \ref{tab:lambda1405_poles}.
  The red line shows the physical region.}
  \label{fig:z-plane}
\end{figure}

We show ${\rm Re}D^{{\bar K}N\pi\Sigma}$, ${\rm Im}D^{{\bar K}N\pi\Sigma}$ and $C^{{\bar K}N\pi\Sigma}$ in Figs.\,7 and 8, respectively.
In the present paper we just point out that the pole expansion holds in the two-channel case as in the single-channel case.
More detailed results will be discussed elsewhere.

\begin{figure}[htpb]
  \centering
  \begin{tabular}{cc}
    \begin{minipage}{0.5\hsize}
      \centering
      \includegraphics[width=\linewidth]{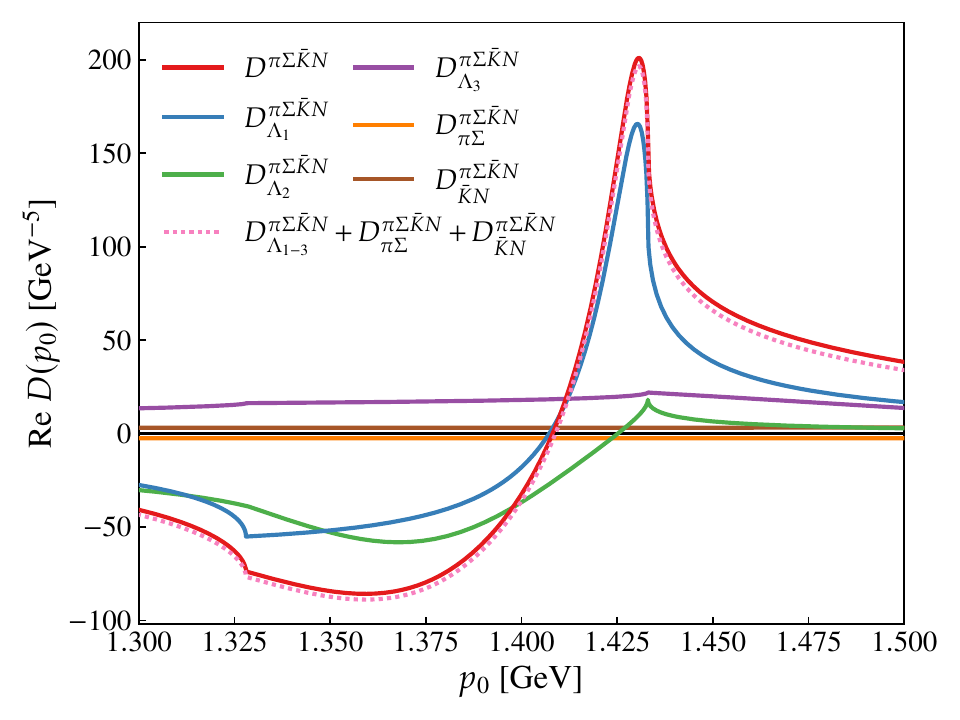}
    \end{minipage}
    \begin{minipage}{0.5\hsize}
      \centering
      \includegraphics[width=\linewidth]{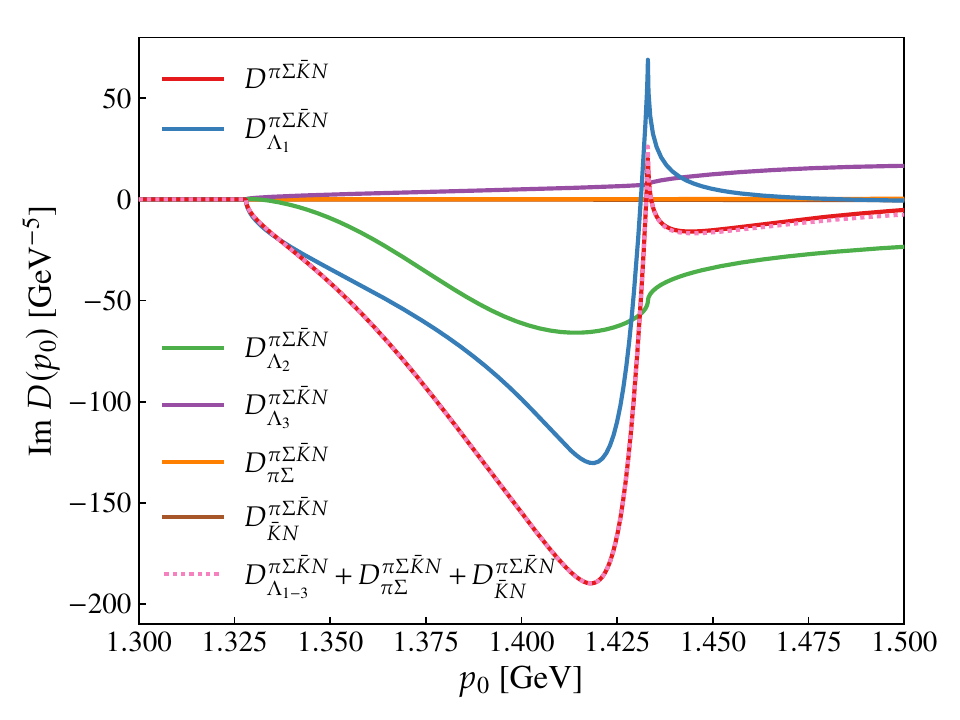}
    \end{minipage}
  \end{tabular}
\caption{$p_0$ dependence of $D^{\bar K N \pi\Sigma}$ and its pole contributions for $|\boldsymbol q|=1$ GeV and $|\boldsymbol q'|=1$ GeV.
The exact (total) $D^{\bar K N \pi\Sigma}$ is shown by the red line. 
The other lines show the contributions from the poles given in Table \ref{tab:lambda1405_poles}.
}
\end{figure}
\begin{figure}[htpb]
  \centering
  \begin{tabular}{c}
    \begin{minipage}{0.5\hsize}
      \centering
      \includegraphics[width=\linewidth]{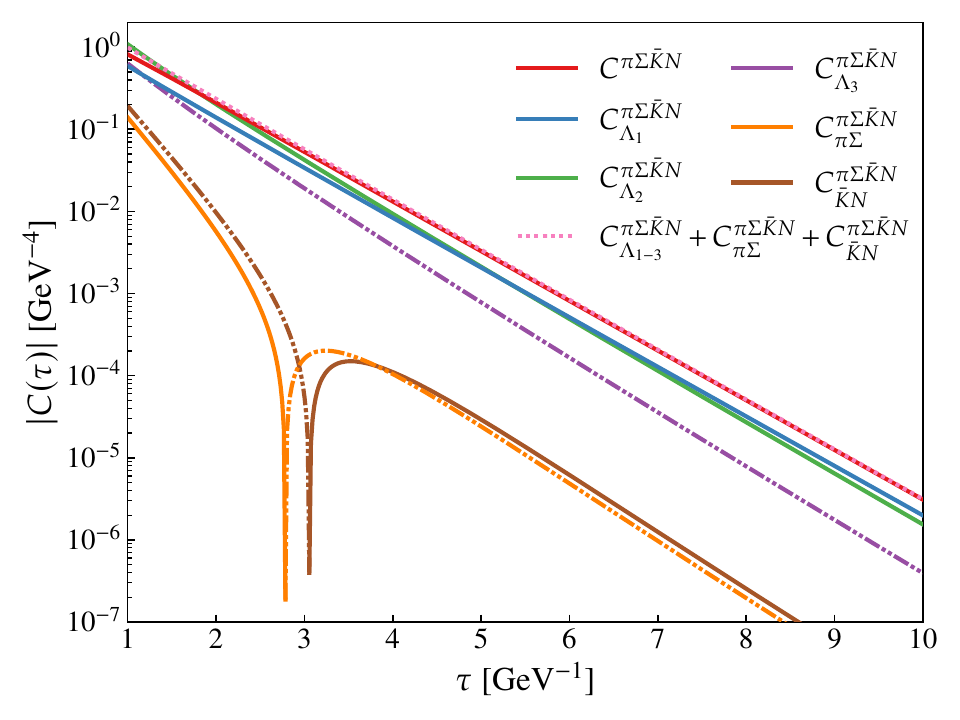}
    \end{minipage}
  \end{tabular}
\caption{$\tau$ dependence of $C^{\bar K N \pi\Sigma}$ and its pole contributions for $|\boldsymbol q|=|\boldsymbol q'|=1$ GeV shown in log scale. 
The dot-dashed lines represent contributions with positive values.
}
\end{figure}

In view of the above we conclude that the pole expansion of the imaginary-time correlation function is valid not only for single-channel but also for coupled-channel (two-channel) scatterings.
It is also suggested that the imaginary-time correlation function can be parametrized by relatively few number of pole positions and residues.
From this observation we propose the pole expansion of the imaginary-time correlation as a method to extract information of unstable states such as masses and widths from the imaginary-time correlation functions obtained by lattice QCD simulations.
The next step is obviously to confirm that we can really extract information of unstable states such as masses and widths from the results of actual lattice QCD simulation.
There is one thing which should be clarified.
Since actual lattice QCD simulation is done with finite volume it is an important question if there is significant finite volume effects in the pole expansion of the two-hadron correlation function or not.
In Ref.\,\cite{Briceno:2017max} argument is given that the finite volume correction of the resonance mass is small in the small width approximation.
However, we would like to confirm it in the context of the pole expansion.

We would like to acknowledge Shoji Hashimoto for discussions and bringing our attention to Ref.\,\cite{Maiani:1990ca}.
Wren Yamada is grateful for the inspiring research environment fostered at iTHEMS.
Koichi Yazaki would like to thank RIKEN iTHEMS and its former director, Tetsuo Hatsuda,
and present director, Satoshi Iso, for allowing me to be its member and the wonderful atmosphere for research.
The work of Osamu Morimatsu and Toru Sato was supported by JSPS KAKENHI Grant Number JP24K07062.
\bibliography{paper}
\end{document}